\documentclass[twocolumn,showpacs,preprintnumbers,amsmath,amssymb,superscriptaddress]{revtex4}

\usepackage{graphicx}
\usepackage{dcolumn}
\usepackage{bm}
\usepackage{subfigure}

\begin{document}

\title{Phase control of spatial interference from two duplicated two-level atoms}

\author{Luling \surname{Jin}}
\affiliation{State Key Laboratory of High Field Laser Physics,
Shanghai Institute of Optics and Fine Mechanics, Chinese Academy of
Sciences, Shanghai 201800, China}

\author{Yueping \surname{Niu}}
\email{niuyp@mail.siom.ac.cn} \affiliation{State Key Laboratory of
High Field Laser Physics, Shanghai Institute of Optics and Fine
Mechanics, Chinese Academy of Sciences, Shanghai 201800, China}

\author{Shangqing \surname{Gong}}
\email{sqgong@mail.siom.ac.cn} \affiliation{State Key Laboratory of
High Field Laser Physics, Shanghai Institute of Optics and Fine
Mechanics, Chinese Academy of Sciences, Shanghai 201800, China}

\date{\today}%

\begin{abstract}
We report the phase control of spatial interference of resonance
fluorescence from two duplicated two-level atoms, driving by two
orthogonally polarized fields. In this closed-loop system, the
relative phase is of crucial importance to the recovery of the
interference patten in the fluorescence light even with strong
driving.
In order to improve the experimental realizability, we propose a
scheme to recover the visibility with fixed relative phase by
adjusting the relative intensities between the two driving fields or
alternatively by using a standing-wave field.
\end{abstract}
%
\maketitle


Young's double-slit experiment is important to our understanding of
the wave nature of light, which exhibits the first-order coherence
properties of light~\cite{scully}. Recently, there has been
considerable interest in the interference of the fluorescence light
from two driven atoms which play the role of the slits in Young's
experiment~\cite{eichmann,int2,int3,int4,int5}. Remarkably, Eichmann
{\it et al}. carried out a very nice experiment where the two slits
were replaced by two $^{198}Hg^+$ ions in a trap and observed the
interference patten in the light scattered from the two
ions~\cite{eichmann}.
However, it has been shown that, in the strong field limit, the
two-particle collective dressed states are uniformly populated, so
that the interference vanishes at strong
driving~\cite{int2,int3,int4,int5}.
This restricts potential applications, e.g., in coherent
backscattering from disordered structures of atoms~\cite{cbs}, the
generation of squeezed coherent light by scattering light off of a
regular structure~\cite{vogel}, the lithography~\cite{double-res},
or precision measurements and optical information processing.

Macovei {\it et al} investigated the radiation from a collective of
atoms~\cite{mm}, and very recently, they proposed a scheme to
recover first-order interference with almost full visibility in
strong fields, by tailoring the surrounding electromagnetic bath
with a suitable frequency dependence, e.g., with the help of
cavities~\cite{mihai}. In the modified reservoirs, the population in
the collective many-particle dressed states was repopulated, so that
the possible scattering pathways was modified and resulted in the
recovery of the interference.

In this Letter, we propose a different scheme to recover the spatial
interference of resonance fluorescence from two duplicated two-level
atoms via controlling the relative phase of driving fields.
The atomic system, investigated before by Bouchene and
coworkers~\cite{Bou1}, is driven by two orthogonally polarized
fields, and thus a closed-loop system is formed, so that the
relative phase becomes very important and can be used to control the
optical property of the medium.
We find that if an appropriate relative phase is chosen, the
interference patten could be recovered even in strong driving
fields, by adjusting the relative intensities between the two
driving fields.
Based on the technology of phase control~\cite{phase}, this scheme
may provide experimental maneuverability. Later, we replace one
driving field with a standing-wave field, and find that the
restriction on the phase shift can be overcome by adjusting the
distance between the atoms and the screen.

\begin{figure}
\begin{center}
\includegraphics[width=7.5cm]{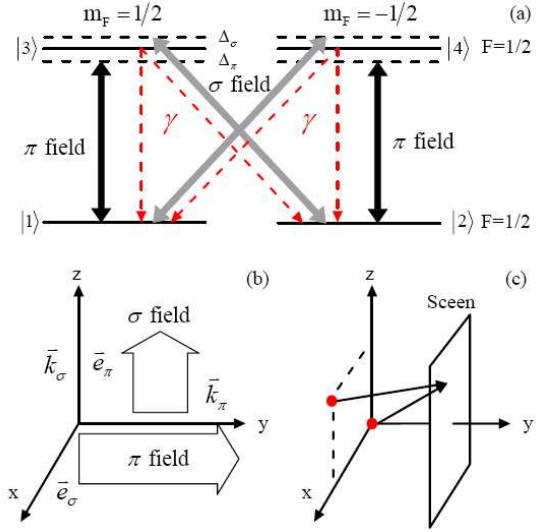}
\end{center}
\caption{\label{fig1}(Color online) (a) Energy level structure for
consideration. The transitions with identical $m_F$ ($ |1\rangle
\leftrightarrow |3\rangle$ and $ |2\rangle \leftrightarrow
|4\rangle$) are coupled by the $\pi -$polarized field, while the
transitions with different $m_F$ ($ |2\rangle \leftrightarrow
|3\rangle$ and $ |1\rangle \leftrightarrow |4\rangle$) are coupled
by $\sigma-$polarized field. (b) Fields configurations. (c)
Schematic diagram of the setup.}
\end{figure}

The atoms used here are modeled as duplicated two-level atoms [see
Fig. \ref{fig1}(a)]. We consider the $F=1/2 \leftrightarrow F=1/2$
transition (energy $\hbar \omega_0$) excited by orthogonally
polarized fields. The system could be realized in $^6 Li$ atom. The
two lower (upper) states $\{|1\rangle,|2\rangle\}$
($\{|3\rangle,|4\rangle\}$ ) with energies $E_1 = E_2$ ($E_3 = E_4$)
correspond to the degenerate states of the lever $^2 S_{1/2}F=1/2$
($^2 P_{1/2}F=1/2$) with $m_F=\pm 1/2$. The transitions with
identical $m_F$ (the transitions $ |1\rangle \leftrightarrow
|3\rangle$ and $ |2\rangle \leftrightarrow |4\rangle$) are coupled
by the $\pi -$polarized field, while the transitions with different
$m_F$ ($ |2\rangle \leftrightarrow |3\rangle$ and $ |1\rangle
\leftrightarrow |4\rangle$) are coupled by $\sigma-$polarized field.
Thus, a closed-loop system is formed, and it allows us to control
optical properties of the medium by the phases of the laser fields.
The electric fields, having the same frequency $\omega$, are $\vec
E_\pi \left( {y,t} \right) = \vec e_z \epsilon _\pi \left( y
\right)e^{ - i\left( {\omega t - ky} \right)} +
{\rm{c}}{\rm{.c}}{\rm{.}}$ and
$\vec E_\sigma  \left( {z,t} \right) = \vec e_x \epsilon _\sigma
\left( z \right)e^{ - i\left( {\omega t - ky} \right)} e^{ - i\phi }
+ {\rm{c}}{\rm{.c}}{\rm{.}}$ , where $\epsilon_i$ is the amplitude
with $i \in \{ \sigma, \pi \}$, $\omega$ is the frequency, $k$ is
the wave vector, and $\phi$ is the relative phase between these
driving fields.
We assume that both excited states have the same decay rate $\gamma$
to the lower levels.

The atomic dipole operator is the sum of atomic raising
$\mu^{\uparrow}$ and lowering $\mu^{\downarrow}$ operators whose
compnents are~\cite{mu}
\begin{subequations}
\label{mu}
\begin{align}
\mu _x^ \downarrow   = \mu \left( {\left| 1 \right\rangle
\left\langle 4 \right| + \left| 2 \right\rangle \left\langle 3
\right|} \right)\hat x ,
\allowdisplaybreaks[2]  \\
\mu _y^ \downarrow   =  - i\mu \left( {\left| 1 \right\rangle
\left\langle 4 \right| - \left| 2 \right\rangle \left\langle 3
\right|} \right)\hat y,
\allowdisplaybreaks[2]  \\
\mu _z^ \downarrow   = \mu \left( {\left| 2 \right\rangle
\left\langle 4 \right| - \left| 1 \right\rangle \left\langle 3
\right|} \right)\hat z ,
\end{align}
\end{subequations}
where $\mu _k^ \downarrow $ is the $k$ component of the atomic
dipole, $\mu$ is the dipole matrix element, and $\hat x $, $\hat y$,
and  $\hat z$ are the usual Cartesian unit vectors. In the
interaction picture, the Hamiltonian of the system in an appropriate
roating frame can be written as
\begin{align}
{\rm H} = \hbar \left( {\begin{array}{*{20}c}
   0 & 0 & {\Omega _\pi  } & { - \Omega _\sigma  e^{ - i\phi } }  \\
   0 & 0 & { - \Omega _\sigma  ^* e^{ - i\phi } } & { - \Omega _\pi  }  \\
   {\Omega _\pi  } & { - \Omega _\sigma  e^{i\phi } } & \Delta  & 0  \\
   { - \Omega _\sigma  ^* e^{i \phi } } & { - \Omega _\pi  } & 0 & \Delta   \\
\end{array}} \right),
\label{HI}
\end{align}
where $\Delta=\omega_0-\omega$ is the detuning, and the Rabi
frequencies are defined as $\Omega _\pi   =\mu \epsilon _\pi/2\hbar$
and $\Omega _\sigma   =\mu \epsilon _\sigma/2\hbar$. The dynamics of
the system can be described using density-matrix approach as
\begin{align}
\dot \rho  =  - \frac{i}{\hbar }\left[ {{\rm H},\rho } \right] +
L[\rho \left( t \right)]. \label{main}
\end{align}
The Liouvillian matrix $L[\rho(t)]$, which describes relaxation by
spontaneous decay, is given by
\begin{align}
L[\rho \left( t \right)] = \left( {\begin{array}{*{20}c}
   {\gamma \left( {\rho _{33}  + \rho _{44} } \right)} & 0 & { - \gamma \rho _{13} } & { - \gamma \rho _{14} }  \\
   0 & {\gamma \left( {\rho _{33}  + \rho _{44} } \right)} & { - \gamma \rho _{23} } & { - \gamma \rho _{24} }  \\
   { - \gamma \rho _{31} } & { - \gamma \rho _{32} } & { - 2\gamma \rho _{33} } & { - 2\gamma \rho _{34} }  \\
   { - \gamma \rho _{41} } & { - \gamma \rho _{42} } & { - 2\gamma \rho _{43} } & { - 2\gamma \rho _{44} }  \\
\end{array}} \right).
\end{align}

We define $n_g = \rho_{11}+\rho_{22}$ and $n_e =
\rho_{33}+\rho_{44}$ as the ground and excited populations, and the
coherences $\rho_{\pi}=\rho_{42}-\rho_{31}$,
$\rho_{\sigma}=\rho_{32}+\rho_{41}$ responsible for the $\pi$- and
$\sigma$-polarized radiated fields, respectively. We solve the
density-matrix equation (\ref{main}) in steady state while
considering the situation that both driving fields are exactly
resonant with corresponding transitions ($\Delta=0$) and all
parameters are dimensionless and normalized by $\gamma$ and we have
\begin{subequations}
\label{sol}
\begin{align}
n_e &= \frac{1}{2}\left[ {1 - \frac{{\Omega _\sigma  ^2  + \Omega
_\pi  ^2 }}{{\Omega _\sigma  ^2  + \Omega _\pi  ^2  + 2\left(
{\Omega _\sigma  ^2  - \Omega _\pi  ^2 } \right)^2 + 4\Omega _\sigma
^2 \Omega _\pi  ^2 \left( {\cos 2\phi  + 1} \right)}}} \right] ,
\\
\rho _\sigma  & = \frac{{\Omega _\sigma  \left[ {\left( {\Omega _\pi
^2  - \Omega _\sigma  ^2 } \right)\sin \phi  + i\left( {\Omega _\pi
^2  + \Omega _\sigma  ^2 } \right)\cos \phi } \right]}}{{\Omega
_\sigma  ^2  + \Omega _\pi  ^2  + 2\left( {\Omega _\sigma  ^2  -
\Omega _\pi  ^2 } \right)^2 + 4\Omega _\sigma  ^2 \Omega _\pi  ^2
\left( {\cos 2\phi  + 1} \right)}},
\\
\rho _\pi   &= \frac{{\Omega _\pi  \left[ {i\left( {\Omega _\pi  ^2
+ \Omega _\sigma  ^2 \cos 2\phi } \right) - \Omega _\sigma  ^2 \sin
2\phi } \right]}}{{\Omega _\sigma  ^2  + \Omega _\pi  ^2  + 2\left(
{\Omega _\sigma  ^2  - \Omega _\pi  ^2 } \right)^2 + 4\Omega _\sigma
^2 \Omega _\pi  ^2 \left( {\cos 2\phi  + 1} \right)}}.
\end{align}
\end{subequations}
From the population and the coherences in the steady state, we find
that the relative phase plays an important role in the atomic
dynamics and should effect the interference patten.

What we are interested in is the far-field interference pattern from
two duplicated two-level atoms. We have calculated the steady-state
solutions for the atomic coherences and populations for the case of
a single atom that interacts with two classical laser light field
linearly polarized along the $x$ axis and $z$ axis, respectively.
Now in our calculation of the far-field interference pattern, we
consider that the separation between the atoms is large enough that
they may be treated independently. The observing screen is placed in
the far-field (large $y$) and oriented in the $x-z$ plane,
illustrated in Fig.~\ref{fig1}(c). At a point $(\tau_1,~\tau_2)$ on
the screen (where $\tau_i$ is the light travel time from the $i$th
atom to the observation point, $i=1,~2$), the intensity of the light
is
\begin{align}
I\left( {\tau _1 ,\tau _2 } \right) \propto \left\langle {E_x ^
\uparrow  E_x ^ \downarrow   + E_z ^ \uparrow  E_z ^ \downarrow  }
\right\rangle , \label{I}
\end{align}
where
\begin{align}
E_k^ \uparrow  \left( {t;\tau _1 ,\tau _2 } \right) \propto e^{ -
i\omega \left( {t - \tau _1 } \right)} u _k^ \uparrow   + e^{ -
i\omega \left( {t - \tau _2 } \right)} \mathcal {U}_k^ \uparrow  ,
\label{E}
\end{align}
for $k\in {x,z}$, $u$ and $ \mathcal {U}$ are the atomic dipoles of
the first and second atoms respectively and $\omega$ is the angular
frequency of the laser light. Since the atoms can be considered as
independent and identical, the intensity of the interference pattern
when all the light is detected is given by
\begin{align}
I\left( {\tau _1 ,\tau _2 } \right) &\propto \left\langle {u _x^
\uparrow  u _x^ \downarrow   +  \mathcal {U}_x^ \uparrow \mathcal
{U}_x^ \downarrow   + u _z^ \uparrow  u _z^ \downarrow + \mathcal
{U}_z^ \uparrow   \mathcal {U}_z^ \downarrow  } \right\rangle
\nonumber \\
&+ \left\langle {u _x^ \uparrow  \mathcal {U}_x^ \downarrow  }
\right\rangle e^{i\omega \left( {\tau _1  - \tau _2 } \right)}  +
\left\langle {u _x^ \downarrow   \mathcal {U}_x^ \uparrow  }
\right\rangle e^{ - i\omega \left( {\tau _1  - \tau _2 } \right)}
\nonumber \\
&+ \left\langle {u _z^ \uparrow   \mathcal {U}_z^ \downarrow  }
\right\rangle e^{i\omega \left( {\tau _1  - \tau _2 } \right)}  +
\left\langle {u _z^ \downarrow   \mathcal {U}_z^ \uparrow  }
\right\rangle e^{ - i\omega \left( {\tau _1  - \tau _2 } \right)}.
\label{I2}
\end{align}
Each component in Eq.~\ref{I2} in the steady state is
\begin{subequations}
\begin{align}
\left\langle {u_x^ \uparrow  u_x^ \downarrow  } \right\rangle _{ss}
&= \left\langle {\mathcal {U}_x^ \uparrow  \mathcal {U}_x^
\downarrow  } \right\rangle _{ss}
\nonumber \\
&\propto \mu ^2 \left\langle {\left( {\left| 4 \right\rangle
\left\langle 1 \right| + \left| 3 \right\rangle \left\langle 2
\right|} \right)\left( {\left| 1 \right\rangle \left\langle 4
\right| + \left| 2 \right\rangle \left\langle 3 \right|} \right)}
\right\rangle_{ss}
\nonumber \\
&= \mu ^2 \left\langle {\left| 4 \right\rangle \left\langle 4
\right| + \left| 3 \right\rangle \left\langle 3 \right|}
\right\rangle_{ss}
\nonumber \\
&= \mu ^2 n_e,
\\
\left\langle {u_z^ \uparrow  u_z^ \downarrow  } \right\rangle _{ss}
&= \left\langle {\mathcal {U}_z^ \uparrow  \mathcal {U}_z^
\downarrow } \right\rangle _{ss}
\nonumber \\
&\propto \mu ^2 \left\langle {\left( {\left| 4 \right\rangle
\left\langle 2 \right| - \left| 3 \right\rangle \left\langle 1
\right|} \right)\left( {\left| 2 \right\rangle \left\langle 4
\right| - \left| 1 \right\rangle \left\langle 3 \right|} \right)}
\right\rangle _{ss}
\nonumber \\
&= \mu ^2 n_e,
\\
\left\langle {u_x^ \uparrow  \mathcal {U}_x^ \downarrow  }
\right\rangle _{ss} &= \left\langle {u_x^ \uparrow  \mathcal {U}_x^
\uparrow  } \right\rangle ^ * _{ss}
\nonumber \\
&\propto \mu ^2 \left\langle {u_x^ \uparrow  } \right\rangle _{ss}
\left\langle {\mathcal {U}_x^ \downarrow  } \right\rangle _{ss}
\nonumber \\
&= \mu ^2 \left\langle {\left( {\left| 4 \right\rangle \left\langle
1 \right| + \left| 3 \right\rangle \left\langle 2 \right|} \right)}
\right\rangle _{ss} \left\langle {\left( {\left| 1 \right\rangle
\left\langle 4 \right| + \left| 2 \right\rangle \left\langle 3
\right|} \right)} \right\rangle _{ss}
\nonumber \\
&= \mu ^2 \rho _\sigma  \rho _\sigma  ^ *,
\\
\left\langle {u_z^ \uparrow \mathcal {U}_z^ \downarrow  }
\right\rangle _{ss} &= \left\langle {u_z^ \uparrow  \mathcal {U}_z^
\uparrow  } \right\rangle ^ * _{ss}
\nonumber \\
&\propto \mu ^2 \left\langle {u_z^ \uparrow  } \right\rangle _{ss}
\left\langle {\mathcal {U}_z^ \downarrow  } \right\rangle _{ss}
\nonumber \\
&= \mu ^2 \left\langle {\left( {\left| 4 \right\rangle \left\langle
2 \right| - \left| 3 \right\rangle \left\langle 1 \right|} \right)}
\right\rangle _{ss} \left\langle {\left( {\left| 2 \right\rangle
\left\langle 4 \right| - \left| 1 \right\rangle \left\langle 3
\right|} \right)} \right\rangle _{ss}
\nonumber \\
&= \mu ^2 \rho _\pi  \rho _\pi ^ *.
\end{align}
\label{uU}
\end{subequations}
%
Unlike the results in Refs.~\cite{int4,int5}, the cross term
$\left\langle {u_x^ \uparrow  \mathcal {U}_x^ \downarrow  }
\right\rangle _{ss}$ and $\left\langle {\mathcal {U}_x^ \uparrow
u_x^ \downarrow  } \right\rangle _{ss}$ contribute to the total
intensity due to the driving of the $\sigma$-polarized field that
$|\rho_\sigma|\neq 0$.
Thus the intensity in Eq. (\ref{I2}) is
\begin{align}
I\left( {\tau _1 ,\tau _2 } \right) \propto 4n_e\{1 + \frac{1}{{2n_e
}}\left( {\rho _\sigma  \rho _\sigma  ^ * + \rho _\pi  \rho _\pi ^ *
} \right)\cos \left[ {\omega \left( {\tau _1  - \tau _2 } \right)}
\right]\},
\end{align}

The visibility of the interference patten is defined as
$V=(I_{max}-I_{min})/(I_{max}+I_{min})$. In our duplicated two-level
atomic system, the visibility can be calculated by using the steady
states solutions (Eq.~\ref{sol})
\begin{align}
V& = \frac{1}{{2n_e }}\left( {\rho _\sigma  \rho _\sigma  ^ *   +
\rho _\pi  \rho _\pi  ^ *  } \right)
\nonumber \\
&=\frac{1}{2}\frac{{\Omega _\sigma  ^2  + \Omega _\pi  ^2 }}{{\Omega
_\sigma  ^2  + \Omega _\pi ^2  + 2\left( {\Omega _\sigma  ^2  -
\Omega _\pi  ^2 } \right)^2 + 4\Omega _\sigma  ^2 \Omega _\pi  ^2
\left( {\cos 2\phi  + 1} \right)}}. \label{V}
\end{align}
Compared with Eq.~\ref{sol}(a), it is easy to see that
\begin{align}
V+n_e=1/2.\label{V+n}
\end{align}
%
We noticed that both of the two components polarized in $x$ and $z$
axis contribute to the total intensity detected on the screen. In
this duplicated two-level atomic system, when the $\pi$- and
$\sigma$-polarized fields are applied simultaneously, the two
components could not be separated and the visibility is always less
than one-half.

From Eq.~\ref{V}, we can see that the interference patten of the
resonance fluorescence from two duplicated two-level atoms is
related to the Rabi frequencies of the driving fields, and
especially to the relative phase $\phi$. $V$ as the function of the
relative phase between these two driving fields reaches its maximum
$V_{max}$ when ${\cos 2\phi  =  - 1}$ ( $\phi  = \frac{\pi }{2} \pm
2n\pi $, $n$ is an arbitrary integer):
\begin{align}
V_{max }  = \frac{1}{2}\frac{{\Omega _\sigma  ^2  + \Omega _\pi ^2
}}{{\Omega _\sigma  ^2  + \Omega _\pi  ^2  + 2\left( {\Omega _\sigma
^2  - \Omega _\pi  ^2 } \right)^2 }}. \label{Vm}
\end{align}
%
From Eq.~\ref{sol}(a), in this case, the excited population reaches
the minimum
\begin{align}
n_{e min}  = \frac{1}{2}\left[ {1 - \frac{{\Omega _\sigma ^2  +
\Omega _\pi  ^2 }}{{\Omega _\sigma  ^2  + \Omega _\pi  ^2  + 2\left(
{\Omega _\sigma  ^2  - \Omega _\pi  ^2 } \right)^2}}} \right].
\label{nm}
\end{align}
%

\begin{figure}
\begin{center}
\includegraphics[width=8.0cm]{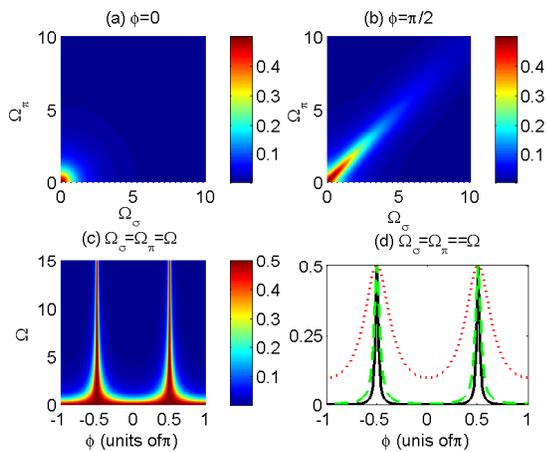}
\end{center}
\caption{\label{fig2}(Color online) The visibility $V$ in different
driving situations: (a) the relative phase is zero, (b) the relative
phase is $\pi/2$, and (c) the intensities of the driving fields are
equal. (d) Some examples in (c). The dotted (red) curve: $\Omega=1$,
the dashed (green) curve: $\Omega=5$, and the solid (black) curve:
$\Omega=10$.}
\end{figure}

It has been confirmed that in strong driving situation, the
interference patten vanishes~\cite{int2,int3,int4,int5}. From our
main results Eqs.~\ref{V} to~\ref{nm}, we find that in our scheme,
the visibility could be realized even in the strong driving
situation due to the relative phase $\phi$.
Figure 2 shows how the visibility evolves under different driving
situations. Without the phase difference, the visibility will fall
towards zero rapidly as the driving field strengths are increased
[see Fig.~\ref{fig2}(a)].
When the phase difference between the two driving fields $\phi$ is
taken into consideration, phase dependent interference of resonance
fluorescence will show up. As we have analyzed from the expression
of $V$, when $\phi=\pi/2$, the visibility will reach its maximum.
From Figs.~\ref{fig2}(b) and (c) we find that even driven by strong
fields, the interference patten could still be observed when the
relative phase is chosen to be around $\pi/2$. The relative phase is
the determinant in the recovery of the interference patten.
Figure~\ref{fig2}(d) shows $V$ vs $\phi$, when the Rabi frequencies
of the two fields are $\Omega_{\sigma}=\Omega_{\pi}=1$, $5$ and $10$
as examples. When the relative phase $\phi$ is exactly equal to
$\pi/2$, we find that $V=1/2$, and from the Eq.~\ref{V+n}, there is
no population on the excited states, that is to say, no fluorescence
could be detected. This can be explained from further investigations
on Eqs.~\ref{Vm} and \ref{nm}. If $\Omega_{\sigma}=\Omega_{\pi}$,
the extremum of $V$ and $n_e$ goes to $V_{max}=1/2$ and $n_{emin}=0$
when $\phi=\pi/2$.

\begin{figure}
\begin{center}
\includegraphics[width=8.0cm]{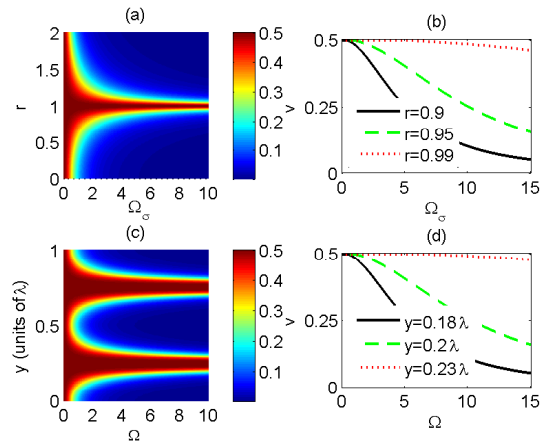}
\end{center}
\caption{\label{fig3}(Color online) The visibility $V$ when
$\phi=\pi/2$. (a) $V$ as the functions of $r$ and the intensity
$\Omega_\sigma$. (b) Some examples in (a) when $r$ is adjusted. (c)
$V$ as the functions of the position $y$ and the intensity $\Omega$.
(d) Some examples in (c).}
\end{figure}

In order to remove the restriction that $\phi$ can not be exactly
$\pi/2$ in experiments when the driving fields have the same
intensities, we investigate the influence of the strengths of those
two driving fields on the interference, especially when
$\phi=\pi/2$. We define the ratio of the Rabi frequencies
$r=\Omega_{\pi}/\Omega_{\sigma}$. It is shown in Fig.~\ref{fig3}(a)
that a peak emerges when $r=1$. As the strengths of the driving
fields increase, the peak becomes narrower. This, however, provides
the feasibility to recover the interference under strong driving
with $\phi = \pi/2$, by choosing the ratio between the two driving
fields properly. We choose $r=0.9$, $0.95$ and $0.99$ for examples
in Fig.~\ref{fig3}(b). It is shown that as $r$ gets closer to $1$,
the visibility could survive even when driven by strong fields.
Thus, by adjusting the relative intensities of driving fields, the
patten would reappearance under strong driving even when
$\phi=\pi/2$.

We notice that the adjusting of the relative intensities works only
when $r$ is modified around one. It is known that a standing-wave
field is periodic in space and oscillates between its minimum and
maximum. An idea came into our mind that we can replace one of the
driving field with a standing-wave field. As interference patten is
observed in the $x-z$ plane, we then use a $\pi$-polarized
standing-wave field, which is applied along the $y$ axis and
therefore $\Omega_{\pi}(y)=\Omega sin(ky)$. The observing screen is
fixed at the end of the cavity, and the atoms are located in $x-z$
plane so that they experience the same driving fields. As the
intensity of the standing wave is position-dependent, the
interference patten in $x-z$ plane is related to the detected
distance between the screen to the plane the atoms are located. In
order to compare with the above work, we choose
$\Omega_{\sigma}=\Omega$, i.e.£¬ $r=sin(ky)$. The result is shown in
Fig.~\ref{fig3}(c). Peaks appear at the antinodes, where $r=1$. By
changing the location of the screen, the visibility could be
recovered. In Fig.~\ref{fig3}(d), we choose the distance $y$ to
correspond with the values of $r$ in Fig.~\ref{fig3}(b), and we
obtain the same results. In other words, when the relative phase is
fixed to $\pi/2$, the interference patten in $x-z$ plane could be
revealed by moving the screen along the $y$ direction. Controlling
the distance between the atoms and the screen is an alternative
choice as compare to adjust the intensities of the driving fields.

In summary, the phase-dependent interference of resonance
fluorescence from two duplicated two-level atoms is investigated.
The interference patten can be recovered in the fluorescence light
of strongly driven atoms due to the relative phase between the two
driving fields. However, when $\phi=\pi/2$, the intensity is zero,
and no fluorescence could be detected. By adjusting the relative
intensities, this problem can be solved. A scheme of recovering the
visibility by using a standing-wave field is proposed too. By
replacing the $\pi$-polarized field with a standing wave, the
interference patten in $x-z$ plane could by revealed by moving the
screen along the $y$ direction.

This work is supported by the 973 Program under Grant No.
2006CB921104, and the National Natural Sciences Foundation of China
(Grant No. 60708008, 60921004, 60978013, 10874194).


\end{document}